# Insights into ultrafast demagnetization in pseudo-gap half metals

Andreas Mann*, Jakob Walowski, Markus Münzenberg[#]

*I. Physikalisches Institut, Georg-August-Universität Göttingen, Germany*

Stefan Maat, Matthew J. Carey, Jeffrey R. Childress,
*San Jose Research Center,
Hitachi Global Storage Technologies, San Jose, CA 95135 USA*

Claudia Mewes
*Department of Physics & Astronomy, University of Alabama, USA*

Daniel Ebke, Volker Drewello, Günter Reiss, Andy Thomas
*Department of Physics, Universität Bielefeld, Germany*

*now at EPF Lausanne, Switzerland
[#]Corresponding author: M. Münzenberg, e-mail: mmuenze@gwdg.de

Interest in femtosecond demagnetization experiments was sparked by Bigot's discovery in 1995. These experiments unveil the elementary mechanisms coupling the electrons' temperature to their spin order. Even though first quantitative models describing ultrafast demagnetization have just been published within the past year, new calculations also suggest alternative mechanisms. Simultaneously, the application of fast demagnetization experiments has been demonstrated to provide key insight into technologically important systems such as high spin polarization metals, and consequently there is broad interest in further understanding the physics of these phenomena. To gain new and relevant insights, we perform ultrafast optical pump-probe experiments to characterize the demagnetization processes of highly spin-polarized magnetic thin films on a femtosecond time scale. The largest spin polarization is obtained in half-metallic ferro- or ferrimagnets, where only one spin channel is populated at the Fermi level, whereas the other one has a gap. This property allows a control of the spin-scattering processes via the electronic structure, and the ultrafast demagnetization is related to the spin-polarization via a Fermi golden rule model. A long demagnetization time correlates with a high spin polarization via suppression of the spin-flip scattering at around the Fermi level. Previous studies have suggested shifting the Fermi energy into the center of the gap by tuning the number of electrons and thereby to study its influence on spin-flip processes. Here we show that choosing isoelectronic Heusler compounds ($Co_2MnSi$, $Co_2MnGe$ and $Co_2FeAl$) allows us to vary the degree of spin polarization between 60% and 86%. We explain this behavior by considering the robustness of the gap against structural disorder. Moreover, we observe that Co-Fe-based pseudo gap materials, such as partially ordered Co-Fe-Ge alloys and also the well-known Co-Fe-B alloys, can reach similar values of the spin polarization. By using the unique features of these metals we vary the number of possible spin-flip channels, which allows us to pinpoint and control the half metals' electronic structure and its influence onto the elementary mechanisms of ultrafast demagnetization.

Since the discovery of ultrafast demagnetization processes on femtosecond time scales, the underlying mechanism is under debate [1,2]. However the last few years have seen the development of the first quantitative models, such as the microscopic three temperature model (M3TM) model [3] and the stochastic Landau-Lifshitz-Bloch (LLB) equations [4,5]. Both models suggest that the spin-scattering $\tau_{max}$ is related to the Gilbert damping $\alpha$ that describes energy dissipation of the magnetic system in quasi equilibrium via the same elementary spin-flip processes [6]. Just recently also progress in the *ab initio* description of Gilbert damping has been made [7], shedding additional light onto a longstanding issue. Taking the experimental values deduced from the Gilbert damping $\alpha$ as the coupling parameter of the magnetic system (magnons) to the electron temperature, the LLB model allows the quantitative description of ultrafast demagnetization versus time without any free parameters [8]. The relation between the Gilbert damping $\alpha$, the spin-scattering at equilibrium, and its connection to ultrafast processes is summarized in a recent review [9]. Nevertheless, new perspectives on the mechanism by alternative theoretical model calculations have reopened the discussion and suggest alternative mechanisms [10, 11, 12]. Here we present investigations of the spin-dependent electronic structure, in particular the half metallic gap size and its relation to the ultrafast energy relaxation. A half metallic gap of 300 meV in one spin channel allows for suppression of spin-flip processes by a factor of 40 compared to the non spin polarized case, but only in a small energy window of <150 meV. A similar suppression of the spin-flip scattering can be achieved in pseudogap materials with non perfect half metallicity over a larger energy window. By engineering the spin dependent electronic structure of the materials, this allows us to control the ultrafast dynamics by a factor of two to four. In the following we will describe how we control the elementary spin-flip processes by design in our experiments.

In a half-metal, only one spin type is present at the Fermi level [13,14]. Consequently, the electron-related properties have the highest possible spin-dependent asymmetry. This property makes these materials interesting for applications in the field of spinelectronics [15] and more recently spincaloritronics [16,17]. Several different classes of half-metallic materials have been investigated: first, some magnetic oxides such as $CrO_2$, $La_{0.66}Sr_{0.33}MnO_3$ and $Fe_3O_4$ exhibit half-metallic behavior. However, even though $CrO_2$ has been shown to have perfect half-metallic electronic structure, it has been difficult to utilize this material in tunneling magnetoresistance (TMR) devices due to the challenging preparation of high-quality films and interfaces [18]. In addition no gradual control of the spin-flip scattering can be realized. Much better suited for experiments, with the possibility of gradual control of the spin polarization, are Heusler compounds, which represent a promising and versatile class of materials. Some ferromagnetic Heusler compounds are half-metallic, others are found to be superconducting [19], and recently, some have been suggested to be a novel class of topological insulators [20]. In the Heusler structure, it is possible to combine different elements to design certain electronic properties. For example, one approach achieving high polarization values is to shift the Fermi energy level into the middle of the bandgap. This can be accomplished by substituting one of the constituents with an element of a different electron count, i.e. to add or to remove electrons [21]. There has been steady progress in using Heusler materials in magnetic tunnel junctions. Thus far, up to 220% magnetoresistance has been observed at room temperature [22]. However, a strong temperature dependence in these Heusler-based tunnel junctions is generally observed, and their TMR ratio typically drops by a factor of two between low and room temperature. This decrease is because half-metallic Heusler compounds ($L2_1$ phase) are very sensitive to structural disorder, giving rise to the formation of defect states deep in the bandgap, as suggested by Picozzi [23]. High spin polarization in the fully ordered $L2_1$ phase in such a compound typically requires high temperature preparation methods to crystallize sufficiently and with the correct chemical order. Nevertheless, partially ordered (e.g., B2 phase) Heusler compounds with a lower ordering temperature also exhibit high spin polarization, in addition to a low structural ordering temperature. By using this wide array of materials we can tune the degree of spin polarization in a sample, e.g. open and reduce the number of spin-flip relaxation channels in non-oxide-based half metals.

Ultrafast pump-probe experiments measure the demagnetization dynamics on a femtosecond time scale [24]. Electrons are photo-excited by a laser pulse (1.5 eV). Efficient thermalization by spin-flip and non-spin-flip electron-electron scattering in the first 10 fs leads to a fast decay of highly energetic electrons that can be observed as a fast, step-like demagnetization. For a half-metal, however, the blocking of spin-flip scattering processes starts to be efficient once all the electrons are relaxed below the energy of the half-metallic gap. Then, the electron and the spin channels are basically isolated, and the spin-flip scattering probability is reduced to zero because no states are available in the other spin channel. The electronic system and the spin system are thermally decoupled, and the demagnetizing time is determined by the weak spin-lattice interaction. Using a Fermi golden rule to model the elementary spin-flip processes, transition rates at the Fermi level from one spin channel to the other are depending on the spin polarization P [25]. This can be related with the demagnetization time $\tau_{el-sp} \sim (1-P)^{-1}$ i.e. the heating rate of the spin-system. It is evident from the preconditions that this model is only valid for gaps larger than the thermal energy. The typical timescale for this process is found to be >100 ps for the oxide half-metals.

For half-metallic Heusler compounds with high spin-polarization, thus far, no evidence of spin-flip blocking has been identified, as might be expected [25,26], opening up questions on the current understanding of mechanisms of spin-flip processes. What can we learn about spin-flip processes and ultrafast magnetization from the seeming failure of this model? Similarly, how do we relate the properties of the Heusler to the spin dependent electronic structure? Previously, it was shown that the magnetization of Heusler compounds follows a Slater-Pauling curve [27]. Thus, it is reasonable to apply a rigid-band picture, where the Fermi level can be shifted simply by changing the number of valence electrons. With that assumption, in a recent study, the ultrafast spin dynamics of $Co_2MnSi$ and $Co_2FeSi$ Heusler compounds were investigated by femtosecond laser excitation. In these materials, the Fermi level is located close to the bottom of the gap for $Co_2MnSi$ and is close to the top of the gap for $Co_2FeSi$. Consequently, states of both bands in $Co_2MnSi$ and $Co_2FeSi$ become thermally populated, even for small excitation energies. The states are not completely blocked and can serve as a channel for fast demagnetization. Simulations reveal that the electron and spin dynamics behave in a semiconductor-like manner: excitation and relaxation channels can be traced in selected bands. For the $Co_2MnSi$ ($Co_2FeSi$) case, a channel of hot holes below (above) the gap dominates the relaxation. In another study, the blocking of spin-flip processes in a series of $Co_2MnAl_xSi_{1-x}$ and $Co_2Fe_XMn_{1-X}Si$ compounds [28,29] was identified by measuring the Gilbert damping using ferromagnetic resonance (FMR). In that study, the Fermi level is shifted from the bottom of the gap for 28 electrons ($Co_2MnAl$) to the top of the gap for 30 electrons ($Co_2FeSi$). A strong dependence of the Gilbert damping with a minimum at $Co_2Fe_{0.4}Mn_{0.6}Si$ (29.4 electrons) was found. Concurrently, a band structure calculation shows that the total density of states at the Fermi level decreases to a minimum. The damping suddenly increases as the Fermi level enters the conduction band, and the half-metallic character breaks down.

In our study, we do not vary the number of electrons, but rather focus on comparing spin-flip processes for the isoelectronic compounds $Co_2MnSi$, $Co_2MnGe$, and $Co_2FeAl$, which all have 29 contributing electrons. The Fermi level should thus be unchanged and located in the gap region. The half-metallic film's growth parameters were optimized using feedback from spin-polarized transport experiments. For the optimized-growth conditions, $Co_2FeAl$ layers were directly grown onto MgO substrates, followed by an MgO tunnel barrier and a Co-Fe counter-electrode with known spin polarization as a reference. The $Co_2FeAl$ layers showed a pronounced B2 order that increased with annealing temperature (Fig. 1a). The integrated intensity of the (400) peak gradually increases, indicating the increase of the quality of the sample with temperature. In the femtosecond demagnetization experiments the time resolved Kerr rotation $\Delta\theta_K$ is measured for $Co_2FeAl$ films that were annealed at various temperatures (Fig. 1b). The pump-probe measurements were carried out using a double-modulated, time-resolved magneto-optic Kerr effect setup [30] and at the same time time-resolved reflectivity of the samples was measured using balanced photo diodes. As a first observation, the demagnetization time $\tau_m$ increases simultaneously with the integrated (400) peak intensity and, thus, with the B2 order. For this $Co_2FeAl$ film, the value of the spin polarization extracted from the TMR measurements with a $Co_{70}Fe_{30}$ counter electrode was 86% (Fig. 1c) [31]. From the Fermi golden rule model $\tau_{el-sp} \sim (1-P)^{-1}$ a slower demagnetization time, which is however still below the picoseconds time range, is expected. One finds an increase from 350+/-5 fs to 385+/-5 fs with annealing temperature.

In the following study, several materials are compared: three isoelectronic Heusler compounds, two CoFe-based materials (one of them the well-known CoFe-B) the half Heusler compound CoMnSb which is also expected to develop half-metallic features, together with standard Ni, Ni-Fe, Co-Fe and $CrO_2$. The demagnetization curves (Fig. 2) are normalized to their demagnetization values at the maximum and shifted for clarity. While Ni's demagnetization is given as a reference at the bottom, the figure shows that the position of the maximum demagnetization $\tau_{max}$ of $Co_2FeAl$ increases to about 1 ps. No step-like signature, specific contribution of non equilibrium electrons to the demagnetization (e.g. for < 100fs), is found [25]. The shift of the maximum demagnetization is small compared to our earlier work on oxide half-metals [11], where a maximum of 84 ps was obtained for $CrO_2$. Here, the previous approach to compare values of $\tau_{max}$ is extended to consider small differences in the polarization that lead to small differences (<10 fs) in $\tau_{max}$. We extract the value of the demagnetization timescale $\tau_m$ using rate-equations based on the three-temperature model by Beaurepaire and coworkers [32]. The coupling strengths between the three temperature reservoirs determine the demagnetization timescale $\tau_m$. Our implementation is based on an analytical solution of this three-temperature model using the sum of two exponential functions [33], related to the two timescales $\tau_m$ and the electron-phonon scattering $\tau_{el\text{-}lat}$. It should be noted that, while we do not consider diffusion processes in the model because we can disregard them in the first few picoseconds, we must fit four parameters (two timescales and two amplitudes). However, one of the parameters, the electron phonon scattering time $\tau_{el\text{-}lat}$ can be eliminated by using reflectivity measurements [34].

In Fig. 3, the demagnetization time $\tau_m$ is plotted against the spin-polarization values of the materials. The values of $\tau_m$ increase from that of Ni at ~160 fs to 380 fs for $Co_2FeAl$, while the spin polarization ranges from 45% for Ni to 86% for $Co_2FeAl$. Notably, the isoelectronic Heusler compounds $Co_2MnSi$, $Co_2FeAl$, $Co_2MnGe$ exhibit spin polarizations P in the range of 61 to 86%. These results fill the gap of intermediate polarization values between standard ferromagnetic metals and near-100% spin-polarized oxides, and display the connection between demagnetization times and spin polarization given by the golden rule model, shown as lines in Fig. 3. For low spin-polarization, the demagnetization timescale $\tau_m$ is dominated by the fast electron-spin relaxation rate $\tau_m \approx \tau_{el\text{-}sp}$ [25]. The intersection of the ordinate for P=0 is determined by the average momentum scattering rate $\tau_{el,0}/c^2$ = 100 fs (c is the spin orbit-coupling mixing parameter). Different values for the anisotropy relaxation rate $\tau_{lat\text{-}sp}$ determine the limiting values at P=1 of $\tau_m \approx \tau_{lat\text{-}sp}$=3 ps and 1 ns for each curve ( the lower red line and upper black line in Fig.3), and are the only free parameters of the model. The demagnetization timescale $\tau_m$ follows the expected evolution indicating the suppression of the spin-flip scattering $\tau_{el-sp} \sim (1-P)^{-1}$ as the dominant factor.

To confirm our results that $Co_2FeAl$ has the highest spin polarization among the full Heusler alloys, we collected additional electronic structure data of the half-metallic films using tunneling spectroscopy. The results, presented in Fig. 4a, show the signature of a half-metallic gap of 300 meV for $Co_2FeAl$, which is absent in the case of $Co_2MnSi$. This gap is also evident by the suppression of the inelastic magnon peak at negative bias voltages (Fig. 4b). Both indicate the half-metallic nature of the material [35]. Therefore, one would naively expect a much larger demagnetization time. If we take $\tau_{el-sp} \sim (1-P)^{-1}$ as the dominating factor in the simple Fermi golden rule model from Müller et al. [25], one can estimate the demagnetization time increase starting from 170 fs as measured for P=0.45 (Ni) using a simple "back-of-the-envelope" calculation: for P=0.86 ($Co_2FeAl$) one expects an increase by a factor of about four to 700 fs, in reasonable agreement with our observed value of 380 fs for $Co_2FeAl$ (Fig. 3).. To further improve the calculation and get more insight into the contributing spin-flip mechanisms, the effect of temperature, together with the small half-metallic gap of 300 meV observed in $Co_2FeAl$, must be taken into account in the model.

In Fig. 3 we used a one-band model that disregarded the electron temperature. This model is also justified in a multi-band situation if the density of states can be described with one averaged polarization value and if the half-metallic gap is larger than the thermal energy of the electrons. In particular, this is the case if the gap is large and the Fermi level is located in the half metallic gap and is separated from the band edges by an energy larger than $k_BT$. Then,

Fermi's golden rule approach for the density of states at the Fermi level can be applied. The spin mixing $c$ can be described by spin-flip transitions at the Fermi level, which are characterized by transition matrix elements (the mixing strength is given by $c = \zeta_{SO}/\Delta E_{ex}$ [11,36], spin-orbit coupling and exchange energy). Spin mixing was first addressed by Elliott [36], who related spin-flip transitions to spin-orbit coupling. The golden rule approach does not necessarily make any prediction concerning the mechanism of the spin-flip scattering and $c$ could be given by any non-spin-conserving process with similar strength replacing the transition probability or microscopic mechanism related to asymmetry of the spin depended density of states for high spin polarization in general. However, the spin-mixing mechanism has been successfully applied for the recent *ab initio* description of the Gilbert damping, at moment the only approach describing its temperature dependent features. Despite these simplifications, the golden rule model has been successful in most cases even though all parameters have so far been treated as being band- and energy-independent.

Clearly, for materials exhibiting only a small gap in one spin channel, the effect of thermal smearing cannot be disregarded. A temperature kernel, given by the derivative of the Fermi function (temperature studied were T=300K, 600 K), is used to describe the effect of temperature on the band structure, depicted in Fig. 5. After folding of the density of states with the temperature kernel, the 300mV-wide gap is significantly blurred due to the thermal excitations. Its value is taken from the tunneling spectroscopy data for $Co_2FeAl$. This demonstrates that, for these small gap materials, the model needs to be extended. To describe the temperature dependence of the magnetic Gilbert damping, Elliott's approach has been extended to a band-resolved method by Kamberský [37,38]. By averaging all possible transitions over the Brillouin zone, including thermal occupation functions, he derived a spin-scattering rate where the temperature-dependent kernel functions are given by $k_{\mu,\upsilon}$ for band index $\mu$ and $\upsilon$, and $|\phi_{\mu,\upsilon}|^2$ is the transition probability.

$$\tau_{el-sp}^{-1} = \left\langle \sum_{\mu,\upsilon} |\phi_{\mu,\upsilon}|^2 k_{\mu,\upsilon} \delta(E_\mu - E_\upsilon - \omega) \right\rangle_{el,lat} \quad (1)$$

$$\approx \frac{c^2}{\tau_{el,0}} \left\langle \sum_{\mu,\upsilon} (1 - P(E)_{\mu,\upsilon})(-\partial_E f_\mu(E)) \delta(E_\mu - E_\upsilon) \right\rangle_{el,lat} \quad (2)$$

In Equation (2) we formulate the transition probability in terms of the band's polarization $P_{\mu,\upsilon}$ and spin mixing $c$ (see details in the Supplementary Information). The temperature-dependent kernels $k_{\mu,\upsilon}$ give rise to intraband (same index) transitions dominating at low temperatures and interband (different index) transitions with spin flips at higher temperatures, whose effect on the ultrafast demagnetization was also discussed by Fähnle in [39] and later reevaluated by Carva [40] explicitly taking into account the electron-phonon and non-equilibrium contributions. The effect is illustrated in Fig. 5 for the specific examples. For $\mu=\nu$ and taking into account only thermal smearing by the electron distribution, the temperature kernel function is given by the derivative of the Fermi distribution $k_\mu = -\partial_E f(E)$, shown in Fig. 5a for T=600 K. The resulting thermally broadened density of states is then calculated by a convolution, and subsequently, the spin polarization P(E) is obtained as explicitly shown for the 1 eV case in Figs. 5a. Three schematic band structures, one with 1 eV and one with 300 meV wide gap, and the third with a pseudo-gap where the density of states is reduced to one-fifth in one spin channel (equal to a moderate spin polarization of P=0.67) over a 1 eV wide gap, are discussed in the following. To compare these three cases, the factor $(1-P)^{-1}$ is calculated in Figs. 5b-d, which is the relevant factor suppressing the spin-flip scattering $\tau_{el-sp}$. The momentum scattering time $\tau_{el,0}$ is defined by the total number of scattering in momentum space (e.g. in an insulator, this goes to zero). At a temperature of 600 K, the suppression factor for spin scattering is $10^4$ in a 1 eV gap, it decreases to 40 for a small gap of 300 meV in a narrow energy window of <150 meV, and it falls off to 10 for a 100 meV gap (not shown). In comparison, for the pseudo-gap case a comparable suppression of a factor of five is found in a wide energy window of 1.1 eV. To derive a model relating the spin relaxation rate to P at the Fermi and the half metallic gap size, we assume an average scattering rate $\tau_{el,0}$ and band independent polarization P. Using the approximations for the Fermi distribution's tail, the expression can be rewritten using the Boltzmann occupation for the electrons and holes yielding

$$\tau_{el-sp} \approx \frac{\tau_{el,0}}{c^2}\left(1-\left(P(E_F)-\frac{(2m^*)^{3/2}}{(2\pi)^2\hbar^3}(k_BT)^{3/2}e^{-E_{gap}/k_BT}\right)\right)^{-1} \quad (3)$$

The factor $(1-P)^{-1}$ is decreased by a Boltzmann probability term that accounts for the thermal excitations. This equation now gives quantitative numbers in the simple model limit for one band taking into account the reduced spin-flip processes in small bandgap materials. We can conclude that, for these Heusler compounds, a gradual transition of the spin-flip probability can be controlled [41] by increasing the half-metallicity (gap width), which is consistent with our observations. However, the thermal effects from the heated electron system with an energy width of 100 meV lead to a very strong reduction of the spin-flip blocking. The reduction can be described by a thermal prefactor taking into account the width of the gap.

The only exception is CoMnSb, which shows the slowest demagnetization time of 18 ps. However, the sample does not have a high spin-polarization value as determined by spin-transport experiments. In that case, an additional effect becomes dominant. This behavior originates from the material's low Curie temperature of only 474 K, which leads to a thermally-induced slowdown of the magnetization's decay. This effect can be demonstrated using either the Landau-Lifshitz-Bloch equation [42] or the microscopic three-temperature model (M3TM) [43,44]. It is a signature of magnetic fluctuations and excited spin waves [45] becoming even more dominant for high spin polarizations, so-called type-II materials [3]. In a fluence-dependent experiment, such a behavior was identified and separated from the typical behavior of a half-metal.

How can we understand the present results? The largest spin polarization in the isoelectronic row of Heusler compounds is found for $Co_2FeAl$, which is striking because it is not expected by band structure calculations. However, its best structural ordering into the B2 phase already at rather low temperatures is beneficial for the increase of the TMR ratio and explains the unexpected superior half-metallic behavior and the increase in the demagnetization time to 380 fs, manifestation of the spin-flip blocking on ultrafast time scales. Similarly, for CoFe-Ge, no perfect half-metallic behavior is predicted, but it exhibits a pseudo-gap around the Fermi level [46,47]. In pseudo-gap materials the same mechanisms as in the Heusler compounds are acting to increase the spin polarization. Adding an sp-atom to the two transition metal's d-states with mixed valence [48] (here the ferromagnets Co and Fe) changes the electronic structure via specific bonds [49]; thus alloying of CoFe with Ge generates gap-like features in the density of states without requiring perfect long range structural order [46,50]. In B2-based alloys this is realized however on a local scale, so that the large spin polarization is robust against disorder in the alloy films. A reduction in the demagnetization time to 310 fs is found. Similarly these strategies can be applied to suppress spin-flip processes at the Fermi level for applications requiring a low magnetic damping or high spin polarization robust against structural disorder. Combinations of materials with Co-Fe-X where the third element X is from the group 13 (Boron group) -16 (chalcogenides) [46,51,52,53,54] open up new possibilities for designing electronic states and magnetic properties. In such pseudo-gap materials the large spin polarization is obtained by a locally induced partial suppression of states in one spin channel. Simultaneously it is possible to block spin-flip processes and thus to suppress spin scattering on ultrafast time scales. Models adding temperature effects allow qualitative prediction of the spin-flip suppression for small-gap half metals.

In summary, we have shown slow demagnetization in femtosecond pump-probe experiments for a series of structurally improved thin-film Heusler compounds with high spin polarization. Taking the best growth conditions for the best materials, the demagnetization time can be gradually increased from 100 fs (for P=0%) to approximately 380 fs (for P=86%) when comparing different materials. The increase can be understood by the factor of $(1-P)^{-1}$ derived from a golden rule approach without considering thermal occupations. Considering the gap size and thermal smearing, we can also describe this increase quantitatively. Our experiments allow us to pinpoint one important mechanism for ultrafast demagnetization directly related to the spin-dependent electronic structure.

**Methods**:
The $Co_2FeAl$ sample has the stacking MgO(001)/ 5 nm MgO/ 20 nm $Co_2FeAl$/ 1.8 nm MgO and was annealed up to 500°C. $Co_2MnSi$ was grown on MgO (001) and has the stacking MgO(001)/ 5 nm MgO/ 15 nm $Co_2MnSi$/ 1.8 nm MgO. The second $Co_2MnSi$ sample was grown on a silicon layer with the stacking Si/ 80 nm Cu/ 5 nm Ta/ 10 nm V/ 15 nm $Co_2MnSi$/ Al (+Ox.). The best samples of the series discussed here were annealed at 400°C for one hour. The CoMnSb sample, a half Heusler compound predicted to show half-metallicity, has a stacking order of Si/ $SiO_2$ / 40nm V/ 100 nm $Co_{32.4}Mn_{33.7}Sb_{33.8}$ / 1.6 nm Al (+$O_x$). All these samples were prepared at the University of Bielefeld.

The $Co_2MnGe$, $(CoFe)_{1-x}Ge_x$, and $Co_{0.5}Fe_{0.5}$ samples were provided by Hitachi GST and were all 25 nm thick films grown on MgO. Although not a Heusler compound, theoretical calculations predicted a robust pseudogap in the minority channel for $(CoFe)_{1-x}Ge_x$.

The Co-Fe-B film was prepared by magnetron sputtering using 2-inch target with composition $Co_{0.2}Fe_{0.6}B_{0.2}$ (analysis Co:Fe 0.32:0.68) in a UHV system with a base pressure of $5\times10^{-10}$ mbar. See detailed preparation method and sample stacking in the Supplementary Materials.

Values for P are taken from Ni [55,56], CoFe [57], Co-Fe-B [51], $Co_2MnSi$ on Si [58], $(CoFe)_{0.72}Ge_{0.28}$ [46], $Co_2FeAl$ [59], CoMnSb (theo.) [60], $CrO_2$ [61] (see table in the Supplementary Information).

**Acknowledgements**:
A.T. and V.D. acknowledge the MIWF of the NRW state government for financial support. M.M. and J.W. acknowledge the funding provided by the German research foundation (DFG) via the SFB 602. A.T., M.M., and G.R. funding by the German research foundation (DFG) via the SpinCaT priority program SPP 1538.

**Author contributions**

A.M. and J.W. performed optical experiments; D.E. and V.D. performed inelastic transport experiments; A.M., J.W, M.M., D.E., V.D., and A.T. analyzed the data; S.M., M.C., and D.E. prepared and characterized the thin films; A.T., S.M. and M.M. designed the research approach; M.M. and C.M. developed the model and theoretical understanding; A.M., M.M., S.M., V.D., and A.T. wrote the manuscript. A.M., M.M., S.M., J.R.C., V.D., G.R. and A.T. and all authors discussed the experiments and the manuscript.

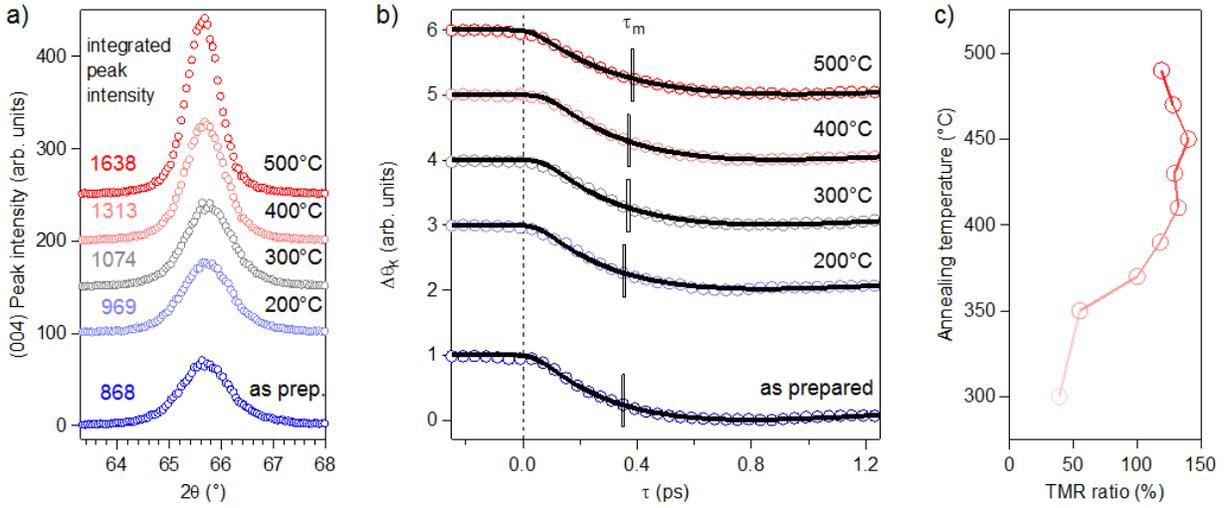

**Fig. 1: Film optimization.** Annealing temperature dependence of $Co_2FeAl$. a) depicts the X-ray diffraction 004-peak, b) shows the demagnetization time (vertical bar marks the value of $\tau_m$), the lines are fits using an analytical function derived from a three-temperature model, c) exhibits the tunnel magnetoresistance ratio (TMR) of $Co_2FeAl$/ MgO/ Co-Fe junctions.

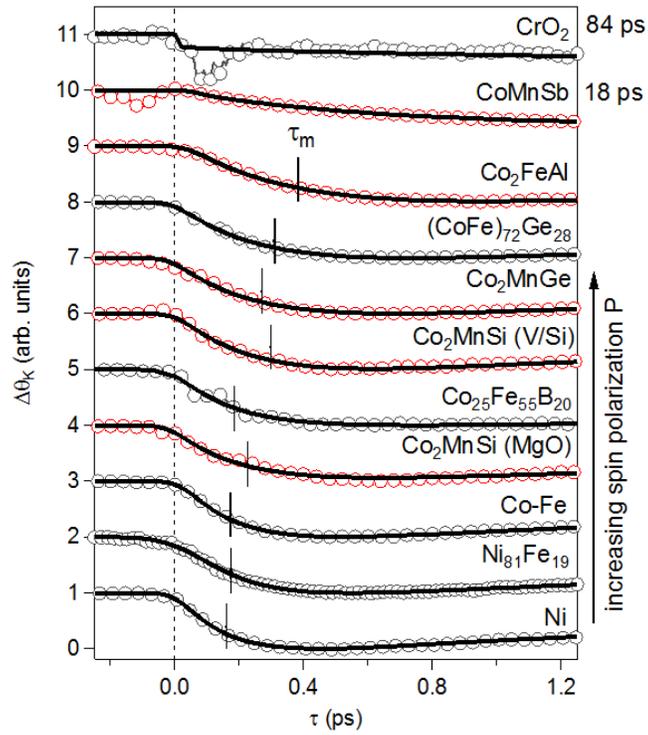

**Fig. 2: Ultrafast demagnetization spectra.** Femtosecond pump-probe experiments for different materials sorted by their spin polarization $P$. Heusler compounds are given in red data points. The lines are fits using an analytical function derived from a three-temperature model and the vertical bar marks the value of $\tau_m$.

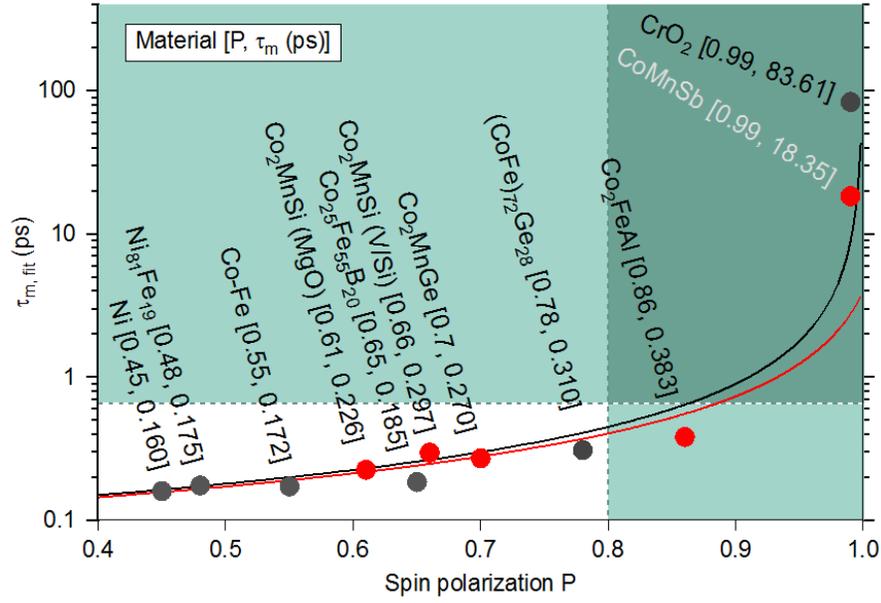

**Fig. 3: Demagnetization time τ$_m$ versus spin polarization P.** The half metallic properties can be classified in a P versus τ$_m$ plot. If the points lie on top of the lines given, spin-flip blocking is the dominant mechanism describing simultaneous increase of P and τ$_m$. The lines are model calculations using Fermi's golden rule approach showing the $\tau_{el-sp} = (1-P)^{-1}$ behavior. For each material the demagnetization time τ$_m$ taken from Fig. 2 is plotted as a function of the spin polarization P. Note the log scale for τ$_m$. Intersections are P=0, $\tau_m = \tau_{el,0}/c^2 = 100\ fs$ and P=1, τ$_m$=3 ps or 1 ns, respectively. Heusler compounds are given in red and corresponding values in brackets (for P and τ$_m$ see legend).

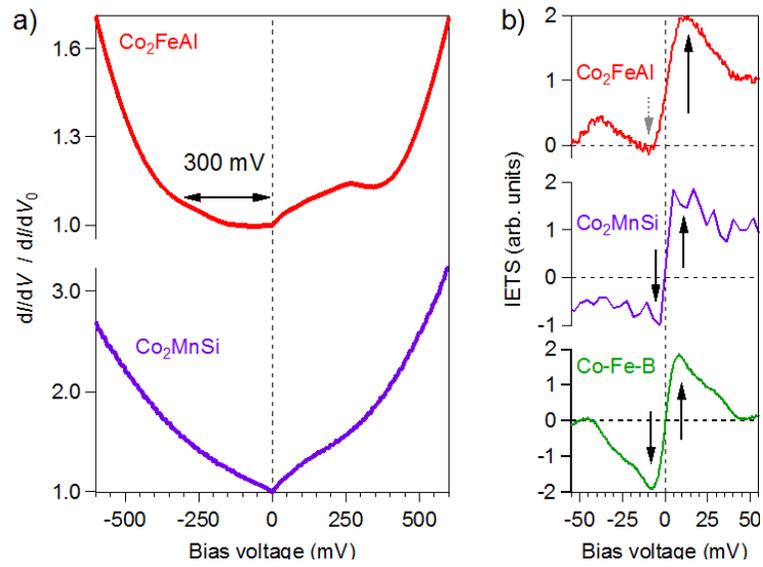

**Fig. 4: Probing the electronic structure.** a) Tunneling spectroscopy of Heusler-based magnetic tunnel junctions. A gap-like feature of 300 mV is visible in the $Co_2FeAl$ spectrum. b) Inelastic electron-tunneling spectra of the Heusler-based and the reference Co-Fe-B junctions. In $Co_2FeAl$, the magnon peak at negative bias is suppressed (compare reference [35]).

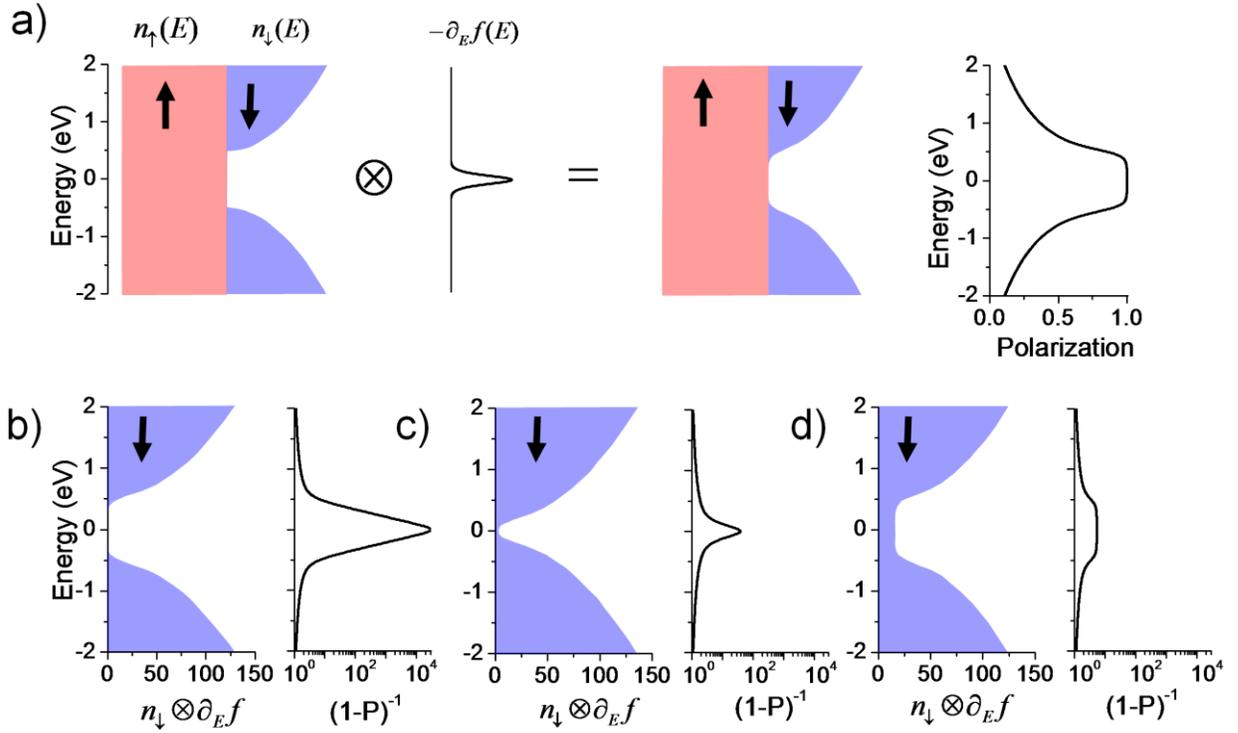

**Fig. 5: Energy dependent golden rule model** a) Model density of states for a one-spin channel with temperature dependent kernel with a width corresponding to T=600K (~60 meV) given by the derivative of the Fermi distribution $-\partial_E f(E)$. The convoluted density of states and the spin polarization P resulting is calculated. In d)-f) the influence of the thermal excitations on the spin-flip suppression factor $\tau_{el-sp} = (1-P)^{-1}$ is calculated in analogy to c) for the three cases: a large gap of 1 eV (b), a small gap of 300 meV (experimental value for $Co_2FeAl$), in the spin down density of states (c), as well as for the pseudogap material, Co-Fe-Ge (d). The suppression factor is given on a logarithmic scale.

# References

[1]  E. Beaurepaire et al., Phys. Rev. Lett. **76**, 4250 (1996).
[2]  A. Kirilyuk, A. V. Kimel, T. Rasing, Rev. Mod. Phys. **82**, 2731 (2010).
[3]  B. Koopmans, G. Malinowski, F. Dalla Longa, D. Steiauf, M. Fähnle, T. Roth, M. Cinchetti, M. Aeschlimann, Nature Mater. **9**, 259 (2010).
[4]  U. Atxitia et al. Appl. Phys. Lett. 91, 232507 (2007).
[5]  I. Radu et al., Nature 472, 205 (2011).
[6]  U. Atxitia and O. Chubykalo-Fesenko, Phys. Rev. B 84, 144414 (2011).
[7]  V. Kamberský, Phys. Rev. B 76, 134416 (2007), K. Gilmore, Y. U. Idzerda and M. D. Stiles, Phys. Rev. Lett. 99, 027204 (2007), A. Brataas, Y. Tserkovnyak and G. E. W. Bauer, Phys. Rev. Lett. 101, 037207 (2008), H. Ebert, S. Mankovsky, D. Ködderitzsch and P. J. Kelly, Phys. Rev. Lett. 107, 066603 (2011).
[8]  U. Atxitia, O. Chubykalo-Fesenko, J. Walowski, A. Mann, and M. Münzenberg, Phys. Rev. B 81, 174401 (2010).
[9]  M. Fähnle, J. Seib, and Ch. Illg, Phys. Rev. B **82**, 144405 (2010), see review M Fähnle and C. Illg J. Phys.: Condens. Matter 23 493201 (2011).
[10] M. Battiato, K. Carva, and P. M. Oppeneer, Phys. Rev. Lett. 105, 027203 (2010).
[11] A. Melnikov, Phys. Rev. Lett. 107, 076601 (2011), M. Wietstruk, et. al., Phys. Rev. Lett. 106, 127401 (2011).
[12] K. Carva, M. Battiato, P.M. Oppeneer, Phys. Rev. Lett. **107**, 207201 (2011), S. Essert and H. C. Schneider, Phys. Rev. B 84, 224405 (2011).
[13] R. A. de Groot, F. M. Möller, P.G. van Engen, and K.H.J. Buschow, Phys. Rev. Lett. **50**, 2024 (1983).
[14] See review by M. Ventkatesan in Handbook of Magnetism and Advanced Magnetic Materials. Editors H. Kronmüller, S.S.P. Parkin, Volume 4: Novel Materials, John Wiley & Sons 2007.
[15] G.-X. Miao, M. Münzenberg and J. S. Moodera, Rep. Prog. Phys. **74**, 036501 (2011).
[16] G. E. W. Bauer, A. H. MacDonald and S. Maekawa, Spin caloritronics, Solid State Commun.**150**, 459–460 (2010).
[17] M. Walter, J. Walowski, V. Zbarsky, M. Münzenberg, M. Schäfers, D. Ebke, G. Reiss, A. Thomas, P. Peretzki, M. Seibt, J.S. Moodera, M. Czerner, M. Bachmann, C. Heiliger, Nature Mater. **10**, 742 (2011).
[18] G.X. Miao, A. Gupta , in *Nanoscale Magnetic Materials and Applications,* J. P. Liu, E. Fullerton, O. Gutfleisch, and D. J. Sellmyer (Eds.), Springer (2009).
[19] B. Balke, G. H. Fecher, J. Winterlik, C. Felser, Appl. Phys. Lett. **90**, 152504 (2007).
[20] S. Chadov,X. Qi,J. Kübler,G. H. Fecher,C. Felser and S. C. Zhang, Nature Mater. **9**, 541 (2010).
[21] T. Graf, S. S. P. Parkin, and C. Felser, Simple Rules for the Understanding of Heusler Compounds, Progress in Solid State Chemistry 39, 1-50 (2011).
[22] S. Tsunegi, Y. Sakuraba, M. Oogane, K. Takanashi, Y.Takanashi,Y. Ando, Appl. Phys. Lett. **93**  (2008) 112506.
[23] S. Picozzi, A. Continenza, and A. J. Freeman, Phys. Rev. B **69**, 094423 (2004).
[24]  A. Kirilyuk, A. V. Kimel, T. Rasing, Rev. Mod. Phys. **82**, 2731 (2010).
[25] G. Müller, J. Walowski, M. Djordjevic, G.X. Miao, A. Gupta, A.V. Ramos, K. Gehrke, V. Moshnyaga, K. Samwer, J. Schmalhorst, A. Thomas, G. Reiss, J.S. Moodera, M. Münzenberg, Nature Mater. **8**, 56 (2009).
[26] D. Steil, S. Alebrand, T. Roth et al., Phys. Rev. Lett. **105**, 217202 (2010)

[27] H.C. Kandpal, G.H. Fecher and C. Felser, J. Phys. D: Appl. Phys. **40**, 1507 (2007)
[28] M. Oogane, T. Kubota, Y. Kota, S. Mizukami, H. Naganuma, A. Sakuma, and Y. Ando, Appl. Phys. Lett. **96**, 252501 (2010).
[29] T. Kubota, S. Tsunegi, M. Oogane, S. Mizukami, T. Miyazaki, H. Naganuma, and Y. Ando, Appl. Phys. Lett **94**, 122504 (2009).
[30] M. Djordjevic, G. Eilers, A. Parge, M. Münzenberg, J. S. Moodera, J. Appl. Phys. **99**, 08F308 (2006)
[31] D. Ebke, P. Thomas, O. Schebaum, M. Schäfers, D. Nissen, V. Drewello, A. Hütten, and A. Thomas, J. Magn. Magn. Mater. **322** (2010) 996.
[32] E. Beaurepaire et al., Phys. Rev. Lett. **76**, 4250 (1996).
[33] F. Dalla Longa, J.T. Kohlhepp, W.J.M. de Jonge, and B. Koopmans, Phys. Rev. B **75**, 224431 (2007).
[34] M. Djordjevic, M. Lüttich, P. Moschkau, P. Guderian, T. Kampfrath, R. G. Ulbrich, M. Münzenberg, W. Felsch, J. S. Moodera, Phys. Stat. Sol. C **3**,1347 (2006).
[35] D. Ebke, V. Drewello, M. Schäfers, G. Reiss, and A. Thomas, Appl. Phys. Lett. **95**, 232510 (2009).
[36] R. J. Elliott, Phys. Rev. **96**, 266 (1954).
[37] V. Kambersky, Phys. Rev. B **76**, 134416 (2007).
[38] K. Gilmore, et al. Phys. Rev. Lett. **99**, 027204 (2007).
[39] M. Fähnle, J. Seib, and Ch. Illg, Phys. Rev. B **82**, 144405 (2010)
[40] K. Carva, M. Battiato, P.M. Oppeneer, Phys. Rev. Lett. **107**, 207201 (2011)
[41] J. Walowski, G. Müller, M. Djordjevic, M. Münzenberg, M. Kläui, C. A. F. Vaz and J. A. C. Bland, Phys. Rev. Lett. **100**, 246803 (2008)
[42] U. Atxitia, O. Chubykalo-Fesenko, J. Walowski, A. Mann, and M. Münzenberg, Phys. Rev. B **81,** 174401 (2010).
[43] B. Koopmans, G. Malinowski, F. Dalla Longa, D. Steiauf, M. Fähnle, T. Roth, M. Cinchetti, M. Aeschlimann, Nature Mater. **9**, 259 (2010).
[44] M. Münzenberg, Nature Mater. **9**, 184 (2010).
[45] M. Djordjevic, M. Münzenberg, Phys. Rev. B **75**, 012404 (2007).
[46] H. Lee, W.H. Butler, T. Mewes, S. Maat, et al. Appl. Phys. Lett. **95**, 082502 (2009).
[47] S. Maat, M.J. Carey, J.R. Childress, Appl. Phys. Lett. **93**, 143505 (2008).
[48] I. Galanakis, P.H. Dederichs, N. Papanikolaou, Phys. Rev. B **66**, 134428 (2002).
[49] W. H. Butler, C.K.A. Mewes, C. Liu, T. Xu, Rational Design of Half-Metallic Heterostructures, arXiv:1103.3855 (2011).
[50] M. Zhu, B.D. Soe, R.D. McMichael, M.J. Carey, S.Maat, and J.R. Childress, Appl. Phys. Lett. **98**, 072510 (2011).
[51] S.X. Huang, T.Y.Chen, and C. L. Chien, Appl. Phys. Lett. **92**, 242509 (2008).
[52] P.V. Paluskar, et al. Phys. Rev. Lett. **100**, 057205 (2008).
[53] S. Wurmehl, et al., Appl. Phys. Lett. **98**, 012506 (2011).
[54] C. Utfeld, et al., Phys. Rev. Lett. **103**, 226403 (2009).
[55] R. J. Jr. Soulen et al., Science **282**, 8588 (1998).
[56] T. H. Kim, and J. S. Moodera, Phys. Rev. B **69**, 020403 (2004).
[57] J. S. Moodera, J. Nassar, G. Mathon, Annu. Rev. Mater. Sci. **29**, 381 (1999).
[58] J. Schmalhorst, S. Kammerer, G. Reiss, and A. Hütten. Appl. Phys. Lett. **86**, 052501, (2005).

[59] D. Ebke et al., Appl. Phys. Lett. 95, 232510 (2009).
[60] Ph. Mavropulos, I. Galanakis, V. Popescu and P.H. Dederichs, J. Phys.:Condens. Matter **16**, S5759-S5762 (2004).
[61] A. Anguelouch et al., Phys. Rev. B **64**, 180408 (2001).

Supplementary Information to

# Insights into ultrafast demagnetization in pseudo-gap half metals

### A. Table metallic half metals

Table 1:

| Material | [ps] | Polarization |
|---|---|---|
| Ni | 0.160 (10) [1] | 0.45 [2] |
| Py | 0.175 (5) | 0.48 [3] |
| CoFe | 0.172 (5) | 0.55 [4] |
| $Co_{0.25}Fe_{0.55}B_{0.2}$ | 0.185 (6) | 0.65 (am.) [5] |
| $Co_2MnSi$ on MgO(001) | 0.226 (7) | 0.61 [6] |
| $Co_2MnGe$ | 0.270 (4) | 0.60 [7] (~0.70 [8]) |
| $Co_2MnSi$ on Si | 0.297 (5) | 0.66 [9] |
| $(CoFe)_{0.72}Ge_{0.28}$ | 0.310 (4) | 0.78 [10] |
| $Co_2FeAl$ | 0.383 (5) | 0.86 [6] |
| $CrO_2$ | 83.6 (5.0) | 0.99 [11] |
| (CoMnSb) | (18.4) | 1.00 (theor. value) [12] |

### B. Sample structures and layer stacking

20nm Ni/ Si(100) (from [1]), 2nm Au/ 12 nm Py / Si(100) (from [13])are grown by e-beam evaporation, in different UHV systems with a base pressure of below UHV $5\times10^{-10}$ mbar.

The 50 nm Co-Fe-B film was prepared by magnetron sputtering using 2-inch targets with compositions $Co_{0.2}Fe_{0.6}B_{0.2}$ (analysis Co:Fe 0.32:0.68) in a UHV system with a base pressure of $5\times10^{-10}$ mbar, capped with 3 nm Ru/ 50 nm $Co_{0.25}Fe_{0.55}B_{0.2}$/ Si (100). The film was not annealed (amorphous).

The $Co_2FeAl$ samples have the stacking 1.8nm MgO/ 20nm $Co_2FeAl$/ 5nm MgO/ MgO(001) and was annealed up to 500°C. $Co_2MnSi$ was grown on magnesium oxide and has the stacking 1.8nm MgO/ 15nm $Co_2MnSi$/ 5nm MgO/ MgO(001), the second was grown on a silicon layer with the stacking 1.6nm Al (+ox.)/ 15nm $Co_2MnSi$/ 10nm V/ 5nm Ta/ 80nm Cu/ Si(100). The best samples of the series discussed here were annealed at 400°C for one hour. The CoMnSb sample, a half Heusler compound predicted to show half metallicity, has a stacking order of 1.6nm Al (+ox) 100nm $Co_{32.4}Mn_{33.7}Sb_{33.8}$/ 40nm V/ Si + ox./ Si(100). All these samples were prepared at the University of Bielefeld.

The series of $Co_2MnGe$, CoFe-Ge and CoFe samples were provided by Hitachi GST and were grown by sputtering as 25 nm thick films on MgO(100), and glass, respectively. $(CoFe)_{1-x}Ge_x$ is predicted to have a robust pseudogap in the minority channel [14,15]. The samples were prepared by cosputtering from separate CoFe (or $Co_2Mn$) and Ge targets, with varied Ge composition and a layer stacking of 3.5nm Ru/ 25nm CoFe(or $Co_2Mn$)-Ge/ 4nm Cu/ 5nm Ta.

### C. Extension of the Fermi golden rule model

We derived in a Fermi golden rule model to estimate the depression of spin-flip transitions at the Fermi level as function of the spin polarization P for the case where it is the dominant term (high spin polarization) [16]. The spin flip rates are described as follows

$$\frac{d}{dt}\left(n_{e,0}^{\uparrow}(E_F) - n_{e,0}^{\downarrow}(E_F)\right) = -\frac{n_{e,0}^{\uparrow}(E_F) - n_{e,0}^{\downarrow}(E_F)}{\tau_{el-sp}} \quad (1)$$

$$\frac{d}{dt}\left(n_{e,0}^{\uparrow}(E_F) - n_{e,0}^{\downarrow}(E_F)\right) = W_{\uparrow\downarrow} - W_{\downarrow\uparrow} \quad (2)$$

In the spirit of Elliott's equation for spin-mixing [17], we use spin-scattering based on the spin mixing parameter $c$. The spin-flip transition rate increases for strong spin-orbit mixing of states at the Fermi level. One can rewrite for large spin polarization P

$$\tau_{el-sp}^{-1} = \frac{c^2}{\tau_{0,el-sp}}(1-P) \quad (3)$$

In this simple equation the general effect of the suppression of the spin-flip transitions is $\tau_{el-sp} \sim (1-P)^{-1}$. The proportionality factor is determined by the mixing parameter $c^2$ (given by $c = \zeta_{SO}/\Delta E_{ex}$, the spin-orbit coupling and the exchange splitting) over the total number of electron scattering events, the average momentum scatting rate $\tau_{el,0}$. All these values that enter are band- and momentum-averaged values given at the Fermi level. In this simplified ferromagnet the effect of the half metallicity is treated simply as an energy independent feature, strictly valid only for energy gaps larger than any electronic excitations when a high spin polarization P leads to a weak coupling between electrons and lattice. However, more realistically the polarization P is not directly probed at the Fermi energy, but rather in a window of approximately 100 meV around it. Although all parameters are taken as band- and energy-independent, the model has been proven successfully in the case of many oxide half metals [16]. In case of materials having small band gaps or where the Fermi energy is close to the band edges using the simplified equation, we predict the correct trend. But to some degree smaller values of $\tau_{el-sp}$ are found experimentally. They originate from temperature induced effects for small half metallic band gaps and/or a Fermi level close to one of the band edges.

To include temperature-related effects into an extended consideration of high spin polarization effects on the spin scattering rate, we go back to a band and temperature resolved approach. It was developed by Kamberský to calculate the intrinsic Gilbert damping. In this general expression including temperature dependence, one averages over all possible transitions for electron and phonon states $\langle\ \rangle_{el,lat}$. Then the spin flip rate is given as average over all possible transitions $\mu, \upsilon$ described by the interaction element $\phi_{\mu,\upsilon}$. Temperature related effects are taken into account by the temperature kernel $k_{\mu,\upsilon}$.

$$\tau_{el-sp}^{-1}(E) = \left\langle \sum_{\mu,\upsilon} |\phi_{\mu,\upsilon}(E)|^2 k_{\mu,\upsilon}(E) \delta(E_\mu - E_\upsilon - \omega) \right\rangle_{el,lat} \quad (4)$$

Here, the dominant excitations of the spin system we consider are spin-flip excitations. Again one can introduce the Elliott's spin-mixing equation [18] to derive an explicit expression of the transition elements. Without loss of generality we assume that the momentum space of scattering events is mainly determined by the reduced number of spin down channel to scatter into in the half metal by P, and the transition rate is then given by the number of electron states in band μ with $n_{\mu\uparrow}$, the number of hole final states $n_{\upsilon\downarrow}$ in band υ and the degree of mixing as determined by the band mixing parameter c [17]:

$$\tau_{el-sp}^{-1}(E) = \left\langle \sum_{\mu,\upsilon} n_{\upsilon\downarrow}(E) c_{\mu,\upsilon}^2 n_{\mu\uparrow}(E) k_{\mu,\upsilon}(E) \delta(E_\mu - E_\upsilon - \omega) \right\rangle_{el,lat} \quad (5)$$

$$= \left\langle \sum_{\mu,\upsilon} \frac{c_{\mu,\upsilon}^2}{\tau_{el,0(\mu,\upsilon)}} (1 - P_{\mu,\upsilon}(E)) k_{\mu,\upsilon} \delta(E_\mu - E_\upsilon - \omega) \right\rangle_{el,lat} \quad (6)$$

$$= \frac{c^2}{\tau_{el,0}} \left\langle \sum_{\mu,\upsilon} (1 - P(E)_{\mu,\upsilon}) k_{\mu,\upsilon} \delta(E_\mu - E_\upsilon - \omega) \right\rangle_{el,lat} \quad (7)$$

To discuss the general trends again, approximations that entered here are band and momentum averaged values assuming a band independent spin mixing c and average scattering rate $\tau_{el,0}$ and one can approximate the expression as given in Eq. 7. To further evaluate the influence of thermal smearing, and the effect for small band gaps, we take into account the electron distributions given by the Fermi function $f(E)$. The thermal smearing in form of a Bloch state live time broadening arising from the phonons resulting in averaged spectral densities describing each band μ, υ will be disregarded. The thermal kernel $k_{\mu,\upsilon} = -\partial_E f(E)$ is given by the derivative of the Fermi function.

$$\tau_{el-sp}^{-1}(E) = \frac{c^2}{\tau_{el,0}} (1 - P(E))(-\partial_E f_\mu(E)) \quad (8)$$

It is evident by a thermal shortening via additional thermally excited possible pathways. In the main article we discuss this factor as a convolution term in the effective band structure to reevaluated the spin-flip suppression factor. The scattering rate $\tau_{el-sp}$ becomes energy-dependent since the phase space available is energy-dependent and increases above the gap region. As often used in the description of semiconductors one can approximate the thermal distribution in the conduction band by a Boltzmann distribution. Integrating over the energy one gets a simple equation that describes general trends related to the temperature and band gap $E_{gap}$ effecting the demagnetization time on the electron-spin scattering time $\tau_{el-sp}$:

$$\tau_{el-sp} \approx \frac{\tau_{el,0}}{c^2} \left( 1 - \left( P(E_F) - \frac{(2m^*)^{3/2}}{(2\pi)^2 \hbar^3} (k_B T)^{3/2} e^{-E_{gap}/k_B T} \right) \right)^{-1} \quad (9)$$

**References**

[1] U. Atxitia, O. Chubykalo-Fesenko, J. Walowski, A. Mann, M. Münzenberg, Phys. Rev. B **81**, 174401 (2010)

[2] R. J. Jr. Soulen, et al., Science 282, 8588 (1998), T. H. Kim and J. S. Moodera, Phys. Rev. B **69**, 020403 (2004).

[3] G.-X. Miao, M. Münzenberg and J. S. Moodera, Rep. Prog. Phys. **74**, 036501 (2011).

[4] J.S. Moodera, J. Nassar, G. Mathon, Annu. Rev. Mater. Sci. **29**, 381 (1999)

[5] S.X. Huang, T.Y.Chen, and C.L. Chien Appl. Phys. Lett. **92**, 242509 (2008).

[6] D. Ebke et al., Appl. Phys. Lett. 95, 232510 (2009).

[7] B.S.D.Ch. Varaprasad, A. Rajanikath, Y.K. Takahashi, and K. Hono, Appl. Phys. Express **3**, 023002 (2010).

[8] Probably higher in our films, J. A. Katine, W. Chen, B. York, and J. R. Childress, J. Appl. Phys. 109, 093912 (2011).

[9] J. Schmalhorst, S. Kammerer, G. Reiss, and A. Hütten. Appl. Phys.Lett. **86**, 052501, (2005).

[10] S. Maat et al., Appl. Phys. Lett. **93**, 143505 (2008).

[11] A. Anguelouch, et al., Phys. Rev. B **64**, 180408 (2001).

[12] Ph. Mavropulos, I. Galanakis, V. Popescu and P.H. Dederichs, J. Phys.:Condens. Matter 16, S5759-S5762 (2004).

[13] J. Walowski, G. Müller, M. Djordjevic, M. Münzenberg, M. Kläui, C. A. F. Vaz and J. A. C. Bland, Phys. Rev. Lett. 100, 246803 (2008).

[14] S. Maat, M.J. Carey, J.R. Childress, Appl. Phys. Lett. **93**, 143505 (2008).

[15] H.Lee, W.H. Butler, T. Mewes, S. Maat, et al., Appl. Phys. Lett. **95**, 082502 (2009).

[16] G. Müller, J. Walowski, M. Djordjevic, G.X. Miao, A. Gupta, A.V. Ramos, K. Gehrke, V. Moshnyaga, K. Samwer, J. Schmalhorst, A. Thomas, G. Reiss, J.S. Moodera, M. Münzenberg, Nature Mater. **8**, 56 (2009).

[17] R. J. Elliott, Phys. Rev. **96** (1954).

[18] V. Kambersky, Phys. Rev. B **76**, 134416 (2007), K. Gilmore, et al. Phys. Rev. Lett. **99**, 027204 (2007).